\documentclass[12pt]{article}
\usepackage{a4,epsfig}
\begin{document}




\begin{titlepage}

\pagenumbering{arabic}
\vspace*{-1.5cm}
\begin{tabular*}{15.cm}{lc@{\extracolsep{\fill}}r}
&
\\
& &
23 February 2015
\\
&&\\ \hline
\end{tabular*}
\vspace*{2.cm}
\begin{center}
\Large 
{\bf \boldmath
 Lienard-Wiechert potentials for charged tachyons\\
and several remarks on the tachyon Cherenkov radiation} \\
\vspace*{2.cm}
\normalsize { 
   
   {\bf V.F. Perepelitsa}\\
   {\footnotesize ITEP, Moscow            }\\ 
}
\end{center}
\vspace{\fill}
\begin{abstract}
\noindent
The Lienard-Wiechert potentials for charged tachyons are deduced in this note.
They turn out to be twice as large as compared to 
the corresponding potentials for ordinary particles.
Several remarks are made on tachyon Cherenkov radiation and its intensity.
In particular, it follows in a straightforward way from the derivation of the 
Lienard-Wiechert potentials that the tachyon Cherenkov radiation angle obeys 
the same formula as that for ordinary particles. 

\end{abstract}
\vspace{\fill}

\vspace{\fill}
\end{titlepage}






\section{Introduction}
We continue a discussion of the properties of faster-than-light particles 
called tachyons, the particles with spacelike momenta, initiated in 
\cite{ttheor}. But before that we would like to recall several historical 
facts concerning the subject.

The first theoretical arguments for the possibility of the existence of
particles with spacelike momenta can be found in a famous paper by Wigner
in which the classification of unitary irreducible representations (UIR's)
of the Poincar\'{e} group was done for the first time~\cite{wigner1}.
In the 1960's Wigner returned to discuss the UIR's of the Poincar\'{e}
group corresponding to particles with spacelike momenta \cite{wigner2}. He has
shown that quantum mechanical equations corresponding to these UIR's
describe particles with imaginary rest mass moving faster than light.
This coincided in time with the appearance of two pioneering works in
which the hypothesis of faster-than-light particles was formulated explicitly,
accompanied by a kinematic description of them \cite{bds}
and by their quantum field theory \cite{fein}.
The particles were called $tachyons$, from the Greek word
$\tau \alpha \chi \iota \sigma$ meaning $swift$ \cite{fein}.

These propositions immediately encountered strong objections
related to the causality principle. It has been shown in several
papers \cite{newton,roln,parment}, in agreement with an earlier remark by
Einstein \cite{ein} (see also \cite{tolman,moller,bohm}),
that by using tachyons as information carriers one can build a causal loop,
making possible information transfer to the past time of an observer.
This is deduced from the apparent ability of tachyons
to move backward in time, which happens when they have
a negative energy provided by a suitable Lorentz transformation, this property
of tachyons being a consequence of the spacelikeness of their four-momenta.
A consensus was achieved that within the special relativity
faster-than-light signals are incompatible with the principle of causality.

Another important problem related to tachyons was their vacuum instability.
It is a well-known problem which usually appears when considering theoretical
models with a Hamiltonian containing a negative mass-squared term (for an
instructive description of the problem see e.g. \cite{nielsen}). Applied
in a straightforward manner to consideration of faster-than-light particles
it results in a maximum, instead of a minimum, of the Hamiltonian for tachyonic
vacuum fields, and leads to the conclusion that the existence of tachyons
as free particles is not possible.

Fortunately, both problems turned out to be mutually connected and, having
hidden loopholes, they were resolved in the 1970's - 1980's,
as described in detail in \cite{ttheor,pvcaus}.

In brief, the causality problem was resolved by combining the tachyon hypothesis
with modern cosmology, which establishes a preferred reference frame:
so called comoving frame, in which the distribution of matter in the universe,
as well the cosmic background (relic) radiation, are isotropic. This changes
the situation with the causality violation by tachyons drastically, since
the fast tachyons needed for a construction of a causal loop (they are called
transcendent tachyons) are extremely sensitive to the metrics of this frame,
which varies in time due to universe expansion. Therefore
this frame has to be involved when considering the propagation of tachyon
signals through space. These signals turn out to be ordered by retarded
causality in the preferred frame, and after causal ordering is
established in this frame, no causal loops appear in any other frame.

Furthermore, in parallel with the causal ordering of the tachyon propagation
one succeeds in obtaining a stable tachyon vacuum which presents the minimum 
of the field Hamiltonian and appears, in the preferred frame, to be an ensemble
of zero-energy, but finite-momentum, on-mass-shell tachyons propagating
isotropically. The boundaries of this vacuum confine the acausal tachyons.

This note is devoted to the question which arises when discussing properties
of electrically charged tachyons, namely, what would be the Lienard-Wiechert 
potentials for charged tachyons obeying Maxwell equations?
It will be shown that the standard approach to the finding of the 
Lienard-Wiechert potentials for ordinary charged particles, 
e.g. that presented in \cite{ll1,ll2}, is applicable to the charged tachyons.

This approach is based on the consideration of a point-like particle.
Indeed, as was argued in \cite{ttheor}, tachyons, if they exist, are
most probably realizations of the infinite-dimensional unitary irreducible 
representations of the Poincar\'{e} group \footnote{Within the anzatz that
elementary particles should be realizations of unitary irreducible     
representations of the Poincar\'{e} group, the only alternative to the 
infinite-dimensional representations suitable for tachyons is a zero-spin
representation corresponding to a scalar tachyon. But scalar tachyon models
have several serious deficiencies discussed in \cite{ttheor,unita2} and are
unlikely to be realistic models for tachyons.} 
and hence can be considered to be string-like objects, i.e. 
possess a non-zero length $l$. However, due the linearity of the
Maxwell equations one can argue that the applied approach is also valid for 
extended tachyons by using a superposition of solutions found for the 
point-like ones. Remarks concerning the tachyon finite length are given 
below, in footnote 3 and Sect.~3.

To simplify the derivation, it will be carried out in the preferred
reference frame, mentioned above, in which the tachyon behaviour is 
governed by retarded causality \cite{ttheor,pvcaus}. 

\section{ Lienard-Wiechert potentials for tachyons}
\setcounter{equation}{0}
\renewcommand{\theequation}{2.\arabic{equation}}
Guided by \cite{ll1} we start with the equations
for the vector and scalar potentials,~${\bf A}$~and~$\Phi$
\begin{equation}
\triangle{\bf A} - \frac{1}{c^2} \frac{{\partial^2}{\bf A}}{{\partial t^2}}
 = -4 \pi {\bf j},
\end{equation}
\begin{equation}
\triangle\Phi-\frac{1}{c^2} \frac{{\partial^2\Phi}}{{\partial t^2}} =-4\pi \rho,
\end{equation}     
where ${\bf j} = e {\bf v} \delta({\bf R}(t))$, $\rho = e \delta({\bf R}(t))$ 
for a charged tachyon of unit charge $e$ moving with the (constant) 
velocity $v$ and separated by the distance ${\bf R}(t)$ from the observation
point (positioned at the origin of the coordinate frame) at the moment $t$. 
Thus, in particular, for the scalar potential $\Phi$ we have
\begin{equation}
\triangle\Phi - \frac{1}{c^2} \frac{{\partial^2\Phi}}{{\partial t^2}} = 
 -4\pi e \delta({\bf R}(t)).
\end{equation}
Everywhere, except the origin of the coordinate frame, 
$\delta({\bf R}) = 0$, and we have the equation
\begin{equation}
\triangle\Phi - \frac{1}{c^2} \frac{{\partial^2\Phi}}{{\partial t^2}} = 0,
\end{equation}
whose $retarded$ solution has the form \footnote{We omit, as usual, the 
advanced solution of (2.4) having the form $\Phi = \frac{\Psi (t + R/c)}{R}$, 
in accordance with the remark made in the Introduction about  
validity of retarded causality for tachyons in the preferred reference frame.} 
\begin{equation}
\Phi = \frac{\chi (t - R(t)/c)}{R(t)},
\end{equation}
where $\chi$ is an arbitrary function. Its choice is dictated by the need 
to obtain the correct value of the potential near 
${\bf R} = 0$. This can be done by noting that as $R \rightarrow 0$, 
the potential tends to infinity and therefore its derivatives with respect 
to the coordinates increase more rapidly than its time derivative.
Consequently as $R \rightarrow 0$, we can neglect the term 
$(1/c^2)/({\partial^2\Phi}/{\partial t^2})$ compared to $\triangle \Phi$ in 
equation (2.3) \footnote{The neglect of the term
$(1/c^2)/({\partial^2\Phi}/{\partial t^2})$ in the case of fast tachyons 
can be justified by our final consideration of a tachyon to be an extended
(string-like) object with the length $l = l_0 \sqrt{v^2/c^2 -1}$, where $l_0$
is the tachyon intrinsic length \cite{ttheor}. Then the characteristic time of
any process involving fast tachyons cannot be shorter than about $l_0/c$, which
leads, in particular, to final values of ${\partial^2\Phi}/{\partial t^2}$.}. 
Then (2.3) goes into the Poisson equation for a point charge
\begin{equation}
\triangle\Phi =  -4\pi e \delta({\bf R}),~~~R \rightarrow 0.
\end{equation}  
Thus, near ${\bf R} = 0$, relation (2.5) must go to the Coulomb law, from which
it follows that 
\begin{equation}
\Phi = \frac{e \delta(t - R(t)/c)}{R(t)},
\end{equation}
where we have introduced the additional delta function in order to eliminate
the implicit arguments in the function $\chi$.

From this it is easy to get the solution of equation (2.2) for an arbitrary
4-dimensional position of an observer, $P({\bf r}, t)$. Thus, the 
solution for the scalar potential of a tachyon moving in a trajectory
${\bf r_0}(t)$, has the form:
\begin{equation}
\Phi({\bf r}, t)=\int\int\frac{e \delta\Big({\bf r^\prime}-{\bf r_0}(\tau)\Big)}
{|{\bf r - r^\prime}|}~
\delta(\tau -t + \frac{1}{c} |{\bf r - r^\prime}|)~d\tau~dV^\prime ,
\end{equation}
where ${\bf r^\prime} = (x^\prime, y^\prime, z^\prime)$, the volume element 
$dV^\prime = dx^\prime, dy^\prime, dz^\prime$, and $|{\bf r - r^\prime}|$ is
the distance from the volume element $dV^\prime$ to the observation point.
Integrating over $dV^\prime$, we get
\begin{equation}
\Phi({\bf r}, t) = e \int \frac{d\tau}{|{\bf r - r_0}(\tau)|}
\delta\Big(\tau - t + \frac{1}{c} |{\bf r - r_0}(\tau)| \Big). 
\end{equation}
The $\tau$ integration will be done using the formula
\begin{equation}
\delta \Big( F(\tau) \Big)=\sum \frac{\delta(\tau-t_{em})}{|F^\prime(t_{em})|},
\end{equation}
where the sum extends over $t_{em}$ solutions of the equation 

\begin{equation}
 F(t_{em}) = t_{em} -t + \frac{1}{c} |{\bf r - r_0}(t_{em})| = 0.
\end{equation}

In order to find these solutions we first consider the 2-dimensional Minkowski 
diagram in the coordinate plane $x, t$, i.e. the simplified case of 
4-dimensional $r_0 = (x,0,0,t)$ (with $x = vt$), see Fig.~1, with the observer 
position at the origin of the coordinate frame. As can be seen from this 
figure, the tachyon world line intersects the past light cone of the observer, 
symbolized by red ($45^\circ$ and $135^\circ$) lines in Fig.~1, at two world 
points, which results in two solutions of (2.11) for $t_{em}$. Let us denote 
these solutions $t_1$ and $t_2$, as indicated in Fig.~1. The relationship 
between $t_1$ and $t_2$ can be obtained from the equation of proportionality    
of space and time intervals in Fig.~1:
\begin{equation}
\frac{c t_2 + v t_2}{t_0} = \frac{v t_1}{t_1},
\end{equation}
where $t_0$ is the time of tachyon passage through the origin of the $x$ axis.
It can be determined from the definition of the tachyon velocity, e.g. via
equation 
\begin{equation}
v = \frac{c t_1}{t_1 - t_0}
\end{equation}
from which
\begin{equation}
t_0 = \frac{v - c}{v} t_1
\end{equation}
Substituting this expression into (2.12) we get
\begin{equation}
\frac{t_2 (v + c) v} {t_1 (v -c)} = v,
\end{equation}
from which we obtain the relationship between $t_1$ and $t_2$:
\begin{equation}
t_2 = \frac{v -c}{v + c}~t_1.
\end{equation}
\includegraphics[height=12cm,width=20cm]{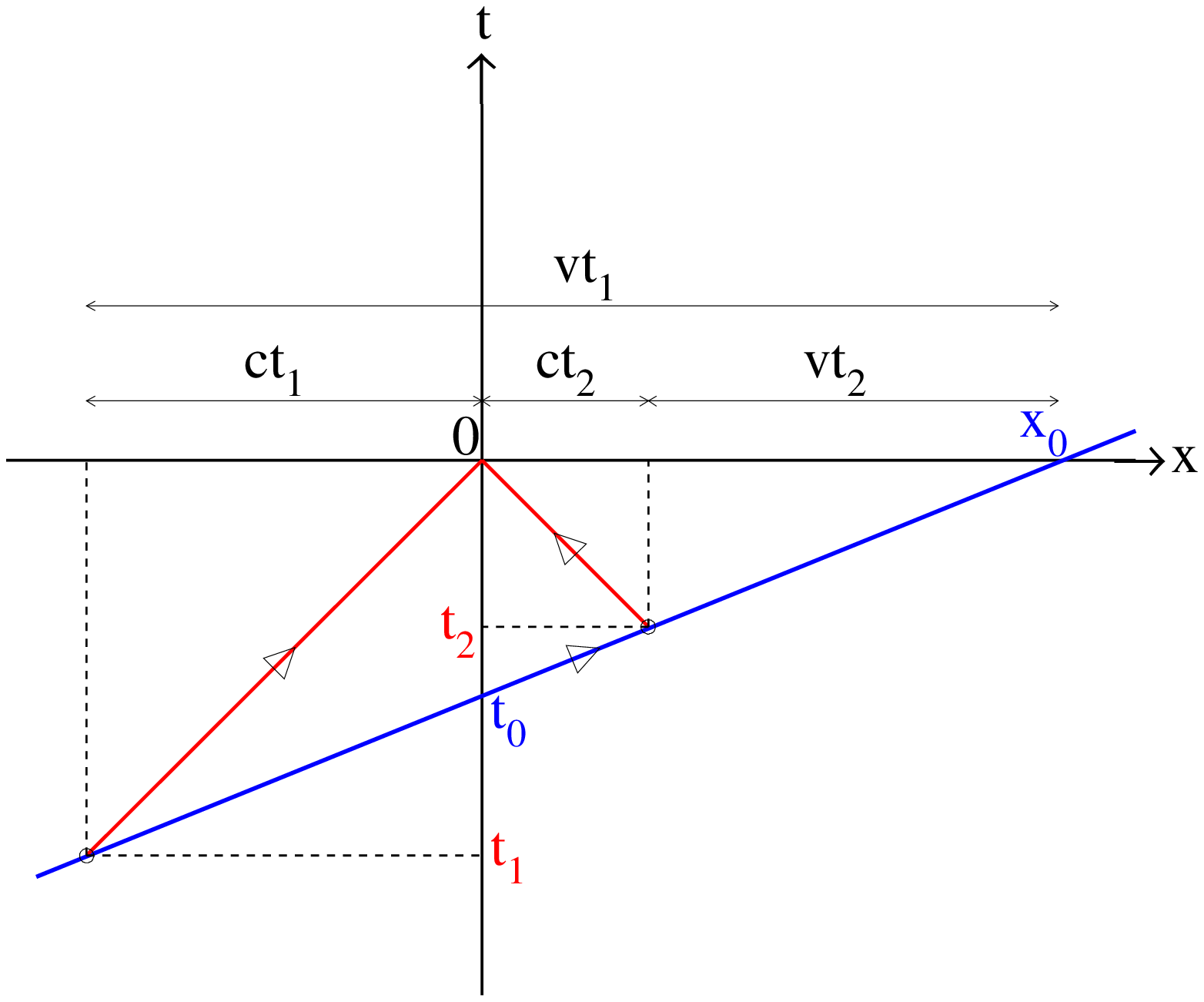}
\vspace{-20mm}
\begin{center} 
{\bf Fig.~1.} The past light cone of the observer and a world line of an 
uniformly moving charged tachyon. $x_0 = v|t_0|$.
\end{center}

Let us consider now the problem in four dimensions. Without loss of 
generality we can hold the position of the observer at the origin of the 
coordinate frame, $x=0$, $y=0$, $z=0$ and $t=0$, with the tachyon velocity 
directed along the $x$ axis as before (thus $x_0 = v|t_0|$), and with the space
coordinates of the tachyon closest approach to the observer being equal to
$0, y, z$. Let $b=\sqrt{y^2+z^2}$, the tachyon impact parameter. 
Then the equation (2.11) in our coordinate frame can be rewritten as 
\begin{equation}
c^2 t^2_{em} = (x_0 + v t_{em})^2 + b^2,
\end{equation}
where $t_{em}$ can be either $t_1$ or $t_2$ (note, they are both negative
in our construction). This gives the quadratic equation
\begin{equation}
t^2_{em} (v^2 - c^2) + 2 x_0 v t_{em} + x^2_0 + b^2 = 0,
\end{equation}
whose solutions are
\begin{equation}
t_{em} = \frac{-v x_0 \pm \sqrt{c^2 x^2_0 - b^2 (v^2 - c^2)}}{v^2 - c^2}
\end{equation}
with the condition required for $t_{em}$ to be negative. We need to consider 
two cases: 
\begin{itemize}
\item [1.] {\bf $v < c$.} 
The expression under the square root is always positive. The root
value is greater than $c x_0$ and therefore greater than $v x_0$. We have
only one solution for $t_{em}$ (with sign $-$ in front of the square root)
keeping the $t_{em}$ negative. This is in an accordance with statements about
the single solution of equation (2.11) in the derivation of the 
Lienard-Wiechert potentials for ordinary charged particles contained in
textbooks on the classical electrodynamics, e.g. in \cite{ll2,jack1}.
The field advances the moving charge. 
\item [2.] {\bf $v > c$.}
The expression under the square root can be positive or negative.  
The latter case will be considered later, in Sect.~3.
In the former case the square root is real and is smaller than $c x_0$, 
and hence smaller than  $v x_0$. Thus both solutions for $t_{em}$ are valid, 
being negative, and have to be accepted (naturally, if $b = 0$ the relation
(2.16) between them is recovered). They have a non-trivial common 
property of the following nature.
\end{itemize}

In the final derivation of the Lienard-Wiechert potentials we will need
the expression for $| R - {\bf v R}/c|$ coming from (2.10), $R = c t_{em}$. 
Let us express it in terms of $b$ and 
$x_0$, considering the process in the impact parameter plane, 
i.e. omitting the time axis, as drawn in Fig.~2. 

Noting that ${\bf v R}/c$ is the distance $D$ in Fig.~2  and 
$|{\bf v R}/c -R|$ is the distance $d$ in that figure, we write first 
$d^2 = e^2 - a^2$, where $a = \beta R \sin{\theta}$, 
while $\sin{\theta} = b/R$, so $a = \beta b$.
Finally, with $e^2 = b^2 + x^2_0$, we obtain 
\begin{equation}
d^2 = b^2 + x^2_0 - \beta ^2 b^2 = x_0^2 - b^2 (\beta ^2 - 1),
\end{equation}
which is (up to a factor of $1/c^2$) just the expression under the square root 
in (2.19). Since this square root is common for both solutions, 
$t_1$ and $t_2$, the values of $|R - {\bf v R}/c|$ are equal one to another 
in both cases. Now we are prepared to integrate (2.9) obtaining
\begin{equation}
\Phi({\bf r}, t) = \frac{e}{|R_1 - {\bf vR_1}/c|}+\frac{e}{|R_2-{\bf vR_2}/c|}=
\frac{2e}{|R - {\bf vR}/c|},  
\end{equation}
where any of two solutions $R_1,~R_2$ can be used in the last fraction. 
\vspace{-12mm}
\includegraphics[height=12cm,width=21.5cm]{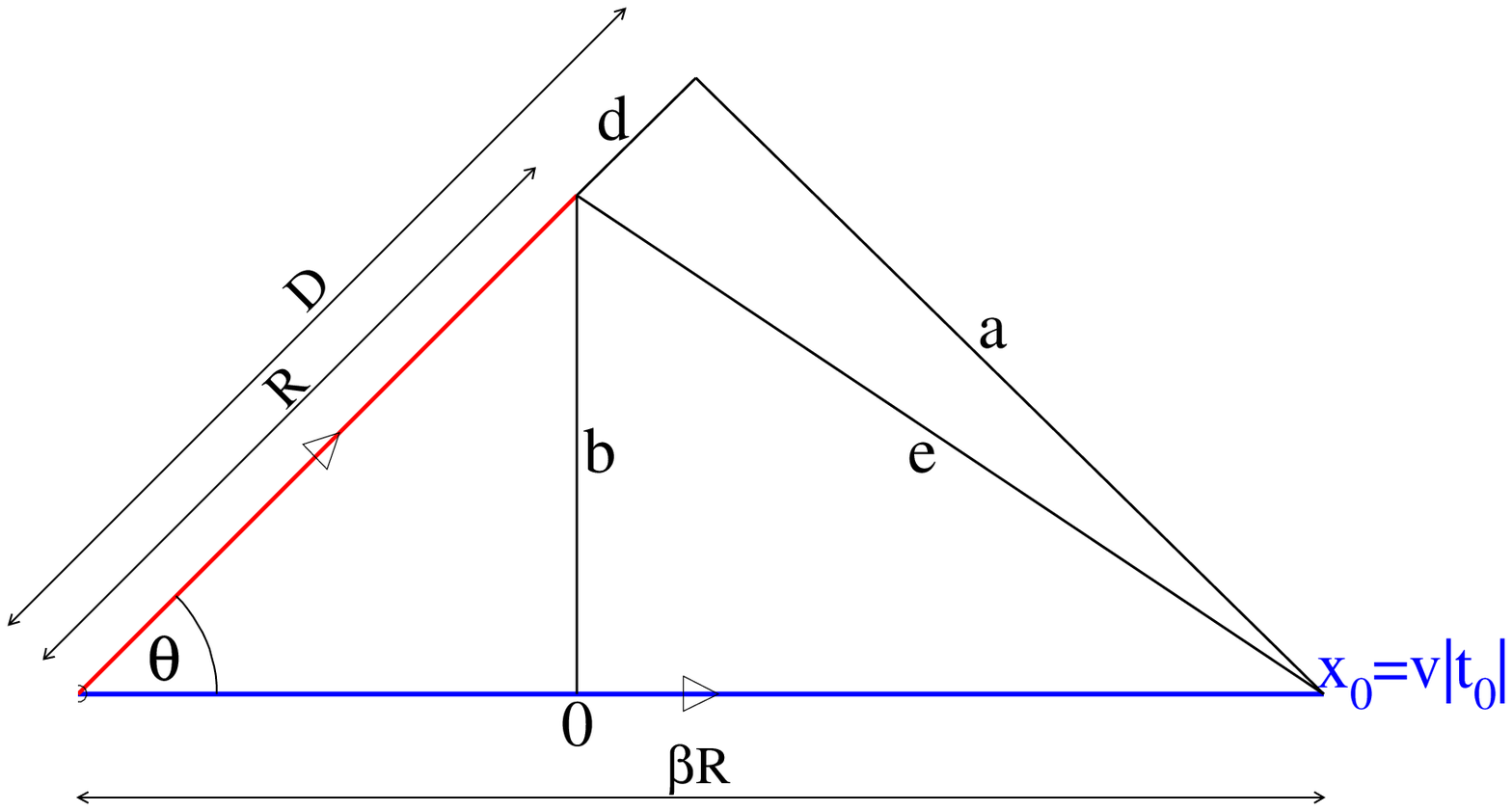}
\begin{center} 
{\bf Fig.~2.} The geometry of retarded potentials. The observer position is 
at the top of line~$b$.
\end{center}
\vspace{5mm}
 
Analogously, for the tachyonic vector potential ${\bf A}$ we obtain
\begin{equation}
{\bf A}({\bf r}, t) = \frac{2 e{\bf v}}{c |R - {\bf vR}/c|}
\end{equation}
These are Lienard-Wiechert potentials for tachyons considered to be point-like
charge carriers. They are twice as large as compared to the corresponding 
potentials for ordinary particles,
\cite{ll2,jack1}:
\begin{equation}
\Phi({\bf r}, t) = \frac{e}{R - {\bf vR}/c},~~~~~~~~~
{\bf A}({\bf r}, t) = \frac{e{\bf v}}{c(R - {\bf vR}/c)} .
\end{equation}
Note that in the both cases the vector potentials $A$ have a single component 
$A_x$.

\section{Remarks on the tachyon Cherenkov radiation in vacuum and the 
characteristic radiation angle}
\setcounter{equation}{0}
\renewcommand{\theequation}{3.\arabic{equation}}
Let us return to the consideration of the expression under the square root in
(2.19). When $c^2 x^2_0 < b^2 (v^2 - c^2)$ the square root is imaginary 
which means the absence of the field in this region. The boundary of this
region, free of the field, with the region where the field is present, is given
by the equation
\begin{equation}
c^2 x^2_0 - b^2 (v^2 - c^2) = 0.
\end{equation} 
This boundary is denoted by the cone $R x_0 S$ in Fig.~3, which 
is the Cherenkov cone 
(this figure reproduces famous figures with wavelets existing in each 
description of the Cherenkov effect in a transparent media, 
e.g. in \cite{jack2}). The cone angle of the Cherenkov radiation in the vacuum 
from a superluminal particle can be easily deduced from Fig.~3:
\begin{equation}
\theta_c = \cos^{-1}(c/v).
\end{equation}
(analogous formula for the tachyon Cherenkov radiation angle in transparent 
media is obtained in Appendix A on the base of a kinematic consideration of
the corresponding reaction of the tachyon Cherenkov radiation).

At the boundary (3.1) the Lienard-Wiechert potentials become infinite as can 
be seen from formulae (2.21), (2.22) taking into account the equation (2.20)
for $d = |R - {\bf v R}/c|$ which vanishes at the boundary.   
This property of the shock wave from a charged particle
moving with faster-than-light velocity in the $vacuum$ was noticed a long time
ago by A.~Sommerfeld who considered, before special relativity  appeared, 
the radiation from an electron moving in vacuum with the superluminal speed
\cite{sommer1,sommer2}. 

\includegraphics[height=12cm,width=20cm]{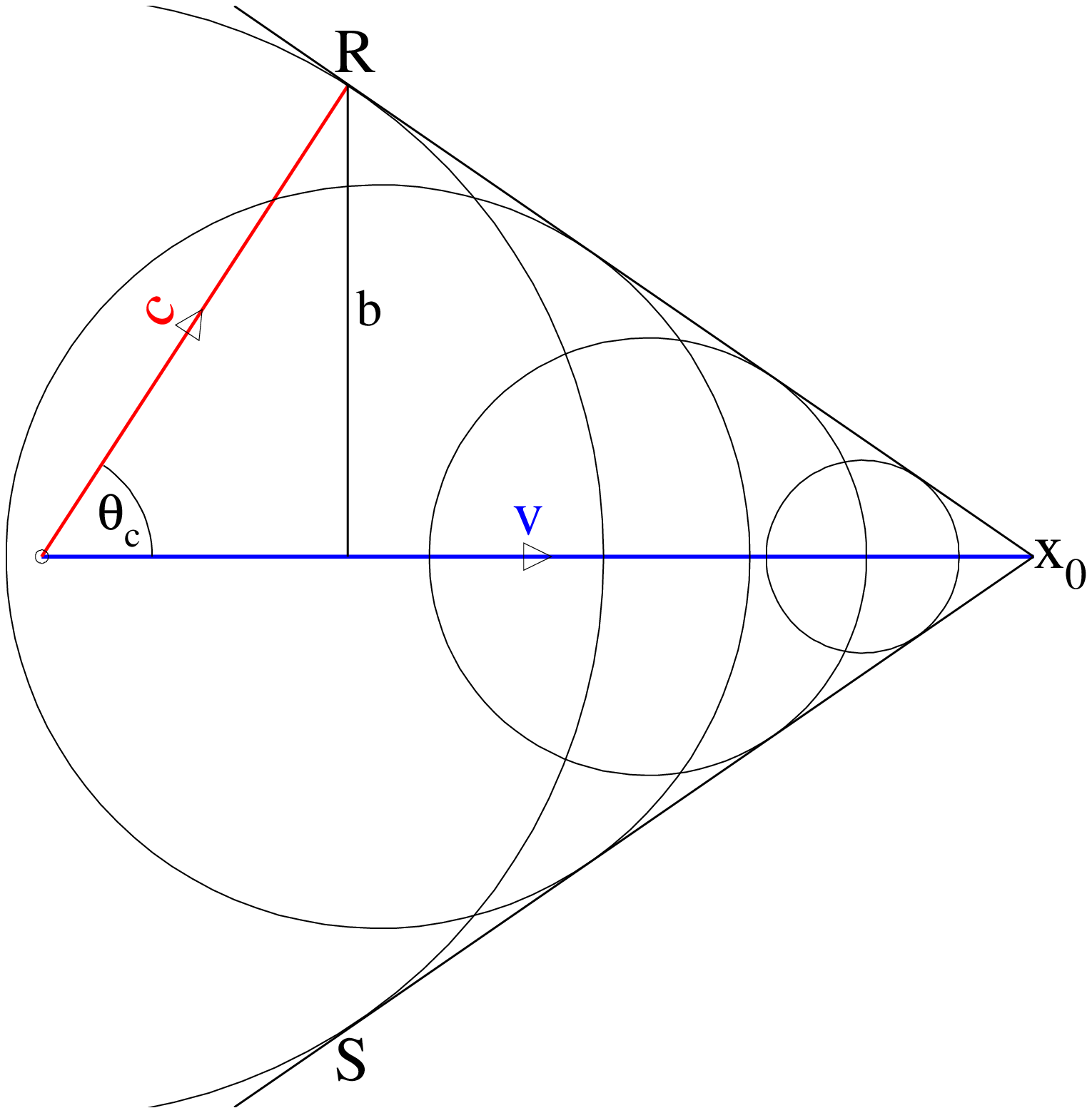}
\begin{center} 
{\bf Fig.~3.} Cherenkov cone of a charged tachyon.
\end{center}

The singularities of potentials
disappear when one considers the faster-than-light particle of a finite size,
which leads to the final depth (spread) of the shock wave front $R x_0 S$ in
Fig.~3. This can be concluded also from the Sommerfeld formula for
energy loss by a superluminal electron due to the radiation in vacuum which
contains a characteristic size of the electron, $a_0$:
\begin{equation}
\frac{dE}{dx} = \frac{9 e^2 (1 - c^2/v^2)}{4  a_0^2},
\end{equation}
It is interesting that the Sommerfeld formula predicts the rate of the
energy loss due to superluminal electron vacuum radiation which is close to the
results of modern calculations for the Cherenkov radiation by a finite-size,
``spherically symmetric" charged tachyon, see \cite{jones,rock,fain}. Note,
it is rather high, but one can expect that such a radiation will be  strongly 
suppressed in a high energy range (for the photon energies comparable with the 
energy of the tachyon) after introduction of a longitudinal tachyon
form-factor \cite{pvcher} (see also Appendix B), and due to quantum selection 
rules \cite{ttheor}. Therefore in our 
derivation of the Lienard-Wiechert potentials we assumed (implicitly) that the
tachyon Cherenkov radiation energy loss is small as compared to the tachyon
energy in order to treat the tachyon velocity $v$ as constant.    

\section{Conclusion}
The Lienard-Wiechert potentials for charged tachyons and some aspects of
the tachyon Cherenkov radiation were considered in this note. It has been
shown that the tachyon Lienard-Wiechert potentials are twice as large as 
compared to the corresponding potentials for ordinary particles. 
The cone angle of the tachyon Cherenkov radiation has been shown to obey the 
same formula as that for ordinary particles, as was expected in \cite{bds}.

\subsection*{Acknowledgements}
\vskip 3 mm
The author thanks Profs. K.~G. Boreskov, F.~S.~Dzheparov, O.~V.~Kancheli for 
stimulating discussions and Dr.~B.~French for a critical reading 
of the manuscript.
\newpage
\section*{Appendix A. Tachyon Cherenkov radiation angle in transparent media}
\setcounter{equation}{0}
\renewcommand{\theequation}{A.\arabic{equation}}
The cone angle of the Cherenkov radiation by tachyons $\theta_c$ in transparent
media is related to the tachyon velocity $v$ and to the radiator refraction 
index $n$ in the same way as for ordinary particles:   
\begin{equation}
\cos \theta_c = \frac{c} {n v}.
\end{equation}
The validity of the formula (A.1) for tachyons follows from the fact that this
formula can be considered as having purely kinematic origin. It can be obtained
from the kinematics of the reaction
\begin{equation}
   t \rightarrow t^\prime + \gamma,
\end{equation}
where $t$ designates a charged tachyon, by use, for example, of 
the equation of four-momentum conservation:  
\begin{equation}
   P = P^\prime + K,
\end{equation}
where $P, P^\prime$ are tachyon four-momenta before and after emission of a
Cherenkov photon, respectively, and $K$ is a four-momentum of the photon.
Moving $K$ to the left side of the equation (A.3) and squaring both sides of it
we get
\begin{equation}
   (P - K)^2 = (P^\prime)^2,
\end{equation}
which reduces to
\begin{equation}
    (P K) = E~\hbar \omega  - \hbar c^2 ({\bf p k}) = 0,    
\end{equation}
where $E$ and $\hbar \omega$ are energies of the initial tachyon and the 
Cherenkov photon, respectively, and ${\bf p}$, $\hbar \bf{k}$ are their 
3-momenta. For photons of optical and near-optical frequencies, propagating 
in medium, the relation between $\omega$ and $k$ is given by \cite{jackson}
\begin{equation}
     \omega ~n(\omega) = c~k(\omega),    
\end{equation}  
where $n(\omega)$ is the refraction index of the medium. Then (A.5) 
transforms to
\begin{equation}
    E  - p c \cos \theta_c ~n(\omega) = 0,    
\end{equation}
from which (A.1) follows, taking into account that tachyon velocity $v$
equals to $pc^2/E$. Naturally, (A.1) reduces to (3.2) for the radiation in
vacuum when $n = 1$.

\section*{Appendix B. The energy loss due to Cherenkov radiation of a
string-like tachyon (the classical approximation)} 
\setcounter{equation}{0}
\renewcommand{\theequation}{B.\arabic{equation}}
In this Appendix we consider the problem of the intensity of the tachyon 
Cherenkov radiation in vacuum for a string-like tachyon of unit electrical
charge in the classical approximation. This approximation is restricted to
the consideration of the tachyon Cherenkov energy loss which is small as
compared to the tachyon energy. In the opposite case a quantum theory of 
tachyon Cherenkov radiation should be developed, 
with the quantum selection rules expected to suppress strongly the radiation. 
In any case, the maximal energy of a single Cherenkov photon is restricted 
in the preferred reference frame (see \cite{ttheor}) by the quantum limit 
$\hbar \omega \leq E_t$, where $E_t$ is the tachyon energy in that frame.

Consider the tachyon as a straight string of a vanishing diameter 
and the Lorentz-contracted (-extended) length of $b = b_0 \sqrt{v^2/c^2-1}$,
as suggested in \cite{ttheor}. Then the distributions of the tachyon current and 
charge entering the Maxwell equations will be defined as
\begin{equation}
{\bf j} = \frac{e {\bf v}}{b}~\theta(x-vt+\frac{b}{2})~
                              \theta(x-vt-\frac{b}{2})~\delta^2(r),
\end{equation}
\begin{equation} 
\rho = \frac{e}{b}~\theta(x-vt+\frac{b}{2})~
                              \theta(x-vt-\frac{b}{2})~\delta^2(r),
\end{equation}
where $v$ is a (constant) velocity of a tachyon moving along the $x$ axis,
and $\theta$'s are Heaviside step functions. 

Similarly to Sect.~2, we start again with the equations
for the vector and scalar potentials,~${\bf A}$~and~$\Phi$
\begin{equation}
\triangle{\bf A} - \frac{1}{c^2} \frac{{\partial^2}{\bf A}}{{\partial t^2}}
 = \frac{-4 \pi e~{\bf v}}{b c}~\theta(x-vt+\frac{b}{2})~
                              \theta(x-vt-\frac{b}{2})~\delta^2(r),
\end{equation}
\begin{equation}
\triangle\Phi-\frac{1}{c^2} \frac{{\partial^2\Phi}}{{\partial t^2}} =
-4\pi~\frac{e}{b}~\theta(x-vt+\frac{b}{2})~
                              \theta(x-vt-\frac{b}{2})~\delta^2(r) 
\end{equation}
(cf. the expression 114.6 in \cite{ess}). The corresponding equations for the 
Fourier components of the potentials are
\begin{equation}
k^2~{\bf A_k (t)}+\frac{1}{c^2}~\frac{{\partial^2}{\bf A_k (t)}}{{\partial t^2}}
 = \frac{4\pi e~{\bf v}}{c}~\frac{\sin{(k_x b/2)}}{k_x b/2}
\exp{(-i k_x v t)},
\end{equation}
\begin{equation}
k^2~\phi_k (t) + \frac{1}{c^2}~\frac{{\partial^2}\phi_k (t)}{{\partial t^2}} =
4\pi e~      \frac{\sin{(k_x b/2)}}{k_x b/2}
\exp{(-i k_x v t)}
\end{equation}
Here $k^2 = k_x^2 + q^2$. Let us introduce the $\omega = k_x v$. 
One can see from equations (B.5), (B.6) that
${\bf A_k (t)} = {\bf A_k}~\exp{(-i \omega t)}$, 
$\phi_k (t) = \phi_k~\exp{(-i \omega t)}$. With this the solutions of these
equations are
\begin{equation}
{\bf A_k (t)}=\frac{4\pi e~{\bf v}}{c}~\frac{\sin{(\omega b/2v)}}{\omega b/2v}~ 
\frac{\exp{(-i\omega t)}}{k^2 - \omega^2/c^2},
\end{equation}
\begin{equation}
\phi_k (t) = 4\pi e~\frac{\sin{(\omega b/2v)}}{\omega b/2v}~
\frac{\exp{(-i\omega t)}}{k^2 - \omega^2/c^2}.
\end{equation}
The Fourier component of the electric field ${\bf E}$ is
\begin{equation}
{\bf E}=4\pi i~e~\Big(\frac{\omega{\bf v}}{c^2}-{\bf k}\Big)~
\frac{\sin{(\omega b/2v)}}{\omega b/2v}~
\frac{\exp{(-i\omega t)}}{k^2 - \omega^2/c^2}.  
\end{equation}
The deceleration force acting on an extended tachyon by its field is
\begin{equation}
{\bf F} = \int {\bf E}(x,r)~\rho(x,r)~dx~d^2 r,
\end{equation}
where ${\bf E}(x,r)$ is the electric field obtained from (B.9) by an inverse
Fourier transform.
\begin{eqnarray}
{\bf F}&=&\frac{4\pi i~e^2}{(2 \pi)^3 b} 
\int \frac{\sin{(\omega b/2v)}}{\omega b/2v}~ 
\frac{\exp{(-i\omega t)}}{k^2 - \omega^2/c^2} 
\Big(\frac{\omega {\bf v}}{c^2} - {\bf k}\Big) dk_x d^2 q  
\int_{vt-b/2}^{vt+b/2} \exp(ik_x x)\exp(iqr) \delta^2(r) dx d^2 r 
\nonumber \\
  &=&\frac{ i~e^2}{2 \pi^2}~
\int \Bigg(\frac{\sin{(\omega b/2v)}}{\omega b/2v}\Bigg)^2~
\frac{\omega {\bf v - k}}{k^2 - \omega^2/c^2}~dk_x d^2 q.
\end{eqnarray}
Integrating (B.11) over the azimuthal angle of the radiation and accounting for
$d k_x = d\omega/ v$, $k^2 = k_x^2 + q^2$, and that the deceleration force
${\bf F}$ is directed along the $x$ axis, one obtains the following expression
for this force:
\begin{equation}
F =\frac{i e^2}{\pi}~\int \Bigg(\frac{\sin{(\omega b/2v)}}{\omega b/2v}\Bigg)^2~
\Bigg(\frac{1}{c^2}-\frac{1}{v^2}\Biggr)~
\frac{q~dq~\omega~d\omega}{q^2 -\omega^2(1/c^2-1/v^2)}.
\end{equation}
Replacing the variable $q$ in this expression by the variable $\xi$ defined as 
$\xi = q^2 -\omega^2(1/c^2-1/v^2)$ one gets
\begin{equation}
F =  \frac{i e^2}{2 \pi}~\Bigg(\frac{1}{c^2}-\frac{1}{v^2}\Biggr)~
\int_{\omega=-\infty}^\infty \int_\xi 
\Bigg(\frac{\sin{(\omega b/2v)}}{\omega b/2v}\Bigg)^2~\frac{d\xi}{\xi}~
\omega~d\omega.
\end{equation}
The integrand of (B.13) contains a pole at $\xi = 0$, corresponding to a
singularity of the potentials on the shock wave front related to the
Cherenkov radiation condition (3.2). Recalling that this singularity can
be avoided by spreading the origin of the shock wave over the tachyon length 
we displace the integration path in the vicinity of the pole 
to the lower half of the $\xi$ complex plane at 
$\omega > 0$ and to the upper half of the plane at $\omega < 0$. Then
the integrals over the real axis of $\xi$ cancel since the integration over
$\omega$ is carried out within symmetric limits while the $\omega$ integrand
is an odd function. Finally, only branching points of the logarithmic function
will give yields to this integral being equal, in sum, to $2 \pi i$. The
result is
\begin{equation}
F = \frac{-e^2}{c^2}~\Bigg(1-\frac{c^2}{v^2}\Biggr)~\int_0^\infty
\Bigg(\frac{\sin{(\omega b/2v)}}{\omega b/2v}\Bigg)^2~\omega~d\omega.
\end{equation}  
In what follows we will use the term of energy loss instead of the deceleration
force, $dE_t/dx = |{\bf F}|$.

To remain within the classical approximation we restrict the integration in 
(B.14) to the $\omega$ values such that $\hbar \omega_{max} << E_t$. Then (B.14)
can be integrated to
\begin{equation}
\frac{dE_t}{dx} = \frac{2 e^2}{b_0^2}~\Bigg[C +\ln{\frac{\omega_{max} b}{v}}~
-~ci\Bigg(\frac{\omega_{max} b}{v}\Bigg)\Bigg],
\end{equation} 
where $C \approx 0.577$ is the Euler constant, and $ci$ denotes the integral
cosine. One can see that the velocity dependence (and, due to (3.2), the 
radiation angle dependence) is rather weak in the final expression for the
stringlike tachyon Cherenkov energy loss.

The analysis of characteristics of the string-like tachyon Cherenkov radiation
spectrum and the consideration of experimental applications of the obtained
results are given in detail in \cite{pvcher}, together with the derivation of 
the Cherenkov energy loss by tachyons with other axially-symmetric form-factors. 
Here we only note the asymptotic behaviour of the energy loss at the tachyon
velocity tending to infinity, i.e. at the tachyon energy $E_t \rightarrow 0$.
The analysis of formula (B.15) shows that at this condition the tachyon
energy loss tends to zero as $E_t^2$.

\end{document}